\documentclass[sigconf]{acmart}
\AtBeginDocument{%
  }

\setcopyright{acmlicensed}
\copyrightyear{2018}
\acmYear{2018}
\acmDOI{XXXXXXX.XXXXXXX}
\acmConference[ACM e-Energy '26]{Make sure to enter the correct
  conference title from your rights confirmation email}{
  2026}{Banff, Canada}
\acmISBN{978-1-4503-XXXX-X/2018/06}

\usepackage{graphicx}
\usepackage{float}
\usepackage{placeins}
\graphicspath{{figures/}}

\begin{document}

\title{Standardized Methods and Recommendations for Green Federated Learning}

\author{Austin Tapp}
\email{atapp@childrensnational.org}
\orcid{0002-0043-6360}
\affiliation{%
  \institution{Children's National Hospital}
  \city{Washington}
  \state{DC}
  \country{USA}
}

\author{Holger R. Roth}
\affiliation{%
  \institution{NVIDIA}
  \city{Santa Clara}
  \state{California}
  \country{USA}}

\author{Ziyue Xu}
\affiliation{%
  \institution{NVIDIA}
  \city{Santa Clara}
  \state{California}
  \country{USA}}

\author{Abhijeet Parida}
\affiliation{%
  \institution{Children's National Hospital}
  \city{Washington}
  \state{DC}
  \country{USA}
}

\author{Hareem Nisar}
\affiliation{%
  \institution{Children's National Hospital}
  \city{Washington}
  \state{DC}
  \country{USA}
}

\author{Marius George Linguraru}
\affiliation{%
  \institution{Children's National Hospital}
  \institution{George Washington University}
  \city{Washington}
  \state{DC}
  \country{USA}
}

\renewcommand{\shortauthors}{Tapp et al.}

\begin{abstract}
Federated learning (FL) enables collaborative model training over privacy-sensitive, distributed data, but its environmental impact is difficult to compare across studies due to inconsistent measurement boundaries and heterogeneous reporting. We present a practical carbon-accounting methodology for FL CO$_2$e tracking using NVIDIA NVFlare and CodeCarbon for explicit, phase-aware tasks (initialization, per-round training, evaluation, and idle/coordination). To capture non-compute effects, we additionally estimate communication emissions from transmitted model-update sizes under a network-configurable energy model. We validate the proposed approach on two representative workloads: CIFAR-10 image classification and retinal optic disk segmentation. In CIFAR-10, controlled client-efficiency scenarios show that system-level slowdowns and coordination effects can contribute meaningfully to carbon footprint under an otherwise fixed FL protocol, increasing total CO$_2$e by $8.34\times$ (medium) and $21.73\times$ (low) relative to the high-efficiency baseline. In retinal segmentation, swapping GPU tiers (H100 vs.\ V100) yields a consistent $\sim$1.7$\times$ runtime gap (290 vs.\ 503 minutes) while producing non-uniform changes in total energy and CO$_2$e across sites, underscoring the need for per-site and per-round reporting. Overall, our results support a standardized carbon accounting method that acts as a prerequisite for reproducible 'green' FL evaluation. Our code is available at \url{https://github.com/Pediatric-Accelerated-Intelligence-Lab/carbon_footprint}.

\end{abstract}

\vspace{-0.8\baselineskip}

\begin{teaserfigure}
  \centering
  \includegraphics[width=0.85\textwidth]{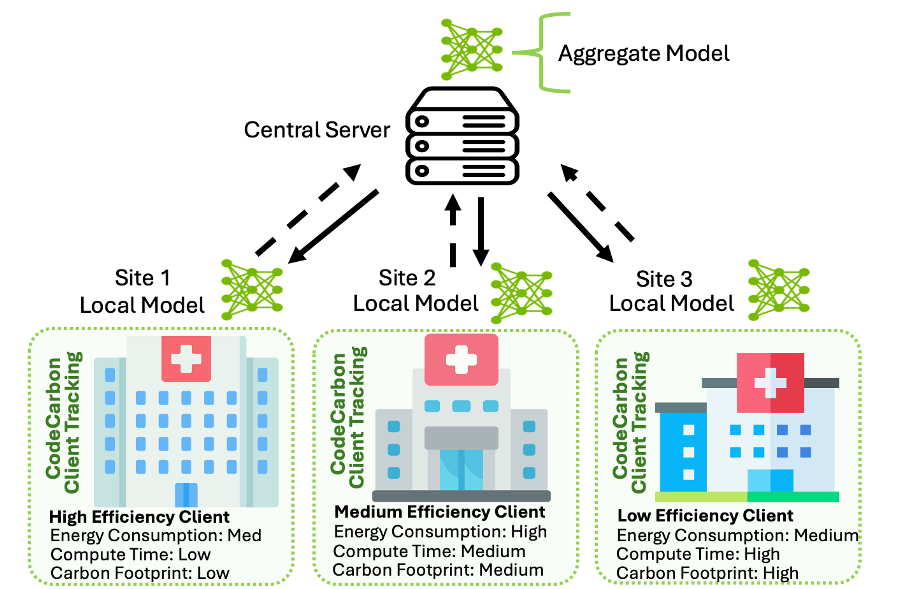}
  \caption{Green Federated Learning simulated through NVFlare, uses a central server to coordinate training and aggregate client updates across three heterogeneous sites, which differ in both data availability and resource efficiency (i.e., efficient GPUs, CPUs, RAM, and energy grids). At each client, CodeCarbon tracks energy use and estimates carbon footprint by accounting for the local grid’s carbon intensity, which reflects how electricity is generated from a regional mix of energy sources with varying emissions. This study illustrates how differences in compute efficiency translate to varying compute time, energy consumption, and emissions—motivating standardized measurement, reporting, and recommendations for green federated learning.}
  \label{fig:teaser}
\end{teaserfigure}

\maketitle

\section{Introduction}

Federated learning (FL) has become a foundational technique for training machine learning models on privacy-sensitive data across multiple distributed sites, offering an appealing solution to preserve security in scenarios where direct data access is typically required. Regardless of deployment scenario, FL is often achieved by iteratively exchanging model updates and aggregating them with algorithms such as Federated Averaging (FedAvg)~\cite{mcmahan2023communicationefficientlearningdeepnetworks}. FL training occurs across many distributed clients whose computing capabilities, utilization patterns, and network links can vary widely. Training pipelines can involve repeated on-device optimization across many participants, and substantial wireless communication with non-trivial coordination overhead all translate into energy consumption and associated carbon emissions. While FL is often motivated by privacy, governance, and bandwidth constraints, the sustainability and environmental implications of FL are increasingly difficult to ignore~\cite{Yang21}. As FL scales to larger models and more heterogeneous deployments, the energy and carbon costs of distributed training become increasingly relevant.

Recently, the ``green AI'' community~\cite{BOLONCANEDO2024128096} has begun to consider sustainability more directly in distributed training~\cite{Qi24}. FL can amplify energy demand on resource-constrained clients because it replaces one centralized training run with many distributed training runs, often repeated across many rounds to converge~\cite{Savazzi22}. Additionally, the communication and coordination patterns of FL (e.g., client selection, straggler handling, and aggregation scheduling) create design choices that directly shape energy use, both on clients and on the server-side orchestration layer~\cite{Ciroi23}. Further, consider that the carbon footprint is not determined solely by how much energy is consumed by FL, but also where and when that energy is consumed~\cite{Mehboob23}. In geographically distributed systems, the same workload can map to substantially different emissions depending on grid mix and regional carbon intensity. Mehboob et al.~\cite{Mehboob23} emphasize that energy's carbon intensity can differ substantially across locations, by up to $60\times$, demonstrating that identical energy usage may lead to widely different carbon outcomes depending on where training occurs.

A growing body of ``green FL'' \cite{Thakur2025GreenFL} research targets lower energy use and improved energy--accuracy tradeoffs through algorithmic and system-level techniques, including energy-aware client scheduling and wireless optimization~\cite{Yang21}, fine-grained gradient compression to reduce communication~\cite{Li21}, resource optimization and device scheduling under energy--performance tradeoffs in edge settings~\cite{Hu22}, and optimized sampling strategies for industrial IoT clients~\cite{Hsu22}. Complementary directions include resource allocation for green edge intelligence~\cite{Salh23}, fairness-aware approaches to curb uneven energy consumption under heterogeneous client behavior~\cite{Albaseer24}, and specialized ``green'' training strategies for lightweight neural networks~\cite{Fontenla24}. In constrained-network settings, quantization and optimized communication designs further reduce energy costs~\cite{Kim24,Wang24}, while cloud--edge--fog orchestration and omnidirectional offloading are explored to minimize end-to-end energy in multi-tier deployments~\cite{Kuswiradyo24}. Alongside these algorithmic advances, recent systems work has begun to formalize measurement and carbon-aware control: Yousefpour et al.~\cite{yousefpour2023greenfederatedlearning} articulate ``Green FL'' and quantify emissions by directly measuring at-scale FL tasks on millions of phones, yielding practical insights into energy efficiency, performance, and time-to-train tradeoffs, while Arputharaj et al.~\cite{arputharaj2025greenfederatedlearningcarbonaware} investigate carbon-aware client selection and training scheduling that leverage slack time and combine $\alpha$-fair carbon allocation with fine-tuning to improve accuracy under carbon constraints using real-world carbon-intensity data. Despite this progress, many studies still quantify energy using domain-specific tasks and non-standardized measurement approaches, which limits cross-study comparability even when reductions in energy consumption or communication overhead are reported.

These efforts collectively underscore an emerging consensus that green FL is a multi-objective systems problem spanning learning dynamics, device constraints, networking, and distributed-system design, and that sustained progress depends on standardized and reproducible carbon-tracking methodologies for both evaluation and mitigation. While prior work demonstrates that carbon accounting in FL is feasible, current approaches remain fragmented and often differ in scope, assumptions, and instrumentation. Although not designed explicitly for FL, CodeCarbon~\cite{mlco2_2025} is widely used as a general-purpose, open-source, carbon emissions tracker that periodically measures or estimates a program's CPU/GPU/RAM energy use and converts it into carbon emissions (CO$_2$e) using location-based grid carbon intensity for logging. In contrast, Qiu et al.~\cite{Qui2023carbonfootprint} estimate client energy by querying GPU utilization via \texttt{nvidia-smi} (with a CPU analogue) and then mapping energy to carbon emissions using client location, but do not rely on a standardized emissions-tracking library. Barbieri et al.~\cite{Barbieri2023} introduce a carbon/energy tracking framework tailored to wireless FL, reporting per-round emissions and device-level CPU/GPU energy, and use these measurements to compare model-parameter compression, sparsification, and quantization strategies across MNIST and CIFAR-10, highlighting trade-offs between learning performance and energy demand. Complementing these measurement frameworks, Cantali et al.~\cite{FedSynthesis2026} extend Flower~\cite{Flower2022} into a carbon-aware FL pipeline by integrating CodeCarbon tracking, MLFlow-based metric logging and visualization, and customized aggregation strategies; on a security-focused FL workflow (GATAKU), they report a 5\% reduction in emissions relative to a standard FL setup. Finally, Feng et al.~\cite{GreenDFL2026} describe an implementation integrated into the open-source DFL system Nebula that estimates emissions from total energy consumption using CPU/GPU power-effectiveness modeling and region-specific carbon intensity; importantly, their accounting also considers communication and aggregation, but relies on a distinct set of modeling assumptions from prior toolchains.

Although these prior works quantify FL carbon footprints, there remains no widely adopted and standardized methodology for estimating emissions and enabling consistent comparisons across FL frameworks, workloads, and deployment settings. Further, no methods integrate seamlessly with production-grade FL orchestrators, offer direct carbon emission estimates through a simulated FL environment, or capture both compute and communication impacts consistently. As a result, carbon footprint reporting remains heterogeneous and often not easily reproducible or verifiable. Even when papers report energy savings or emissions reductions, results can be difficult to reconcile across FL studies because measurements differ in scope, aggregation granularity, and underlying system assumptions.

This paper aims to close these gaps by proposing and evaluating a structured approach for measuring the carbon footprint of FL tasks. The approach is implemented as a lightweight instrumentation layer that integrates CodeCarbon with NVFlare~\cite{NVIDIA25}. It (i) specifies a measurement boundary and logging schema for per-round FL accounting, (ii) records CPU, GPU, and RAM energy via CodeCarbon’s process-level tracking~\cite{mlco2_2025}, (iii) estimates communication-related emissions based on model-update sizes and an explicit networking-energy model, and (iv) offers a configuration-driven workflow to support reproducibility across different hardware and performance settings. We examine the approach using two representative workloads: 1) CIFAR-10 image classification and 2) multi-site retinal optic disk segmentation. We further analyze FL behavior across three client-performance efficiency scenarios (high, medium, and low) to illustrate how idle time and system inefficiencies affect overall emissions. Our results indicate that the proposed methodology provides a practical and transparent approach to studying carbon-footprint characteristics in FL experiments.

This study makes the following contributions:
\vspace{-0.5cm} \begin{itemize}
  \item \textbf{Standardized FL carbon reporting.} We define a minimal set of reporting fields needed for reproducibility.
  \item \textbf{A Simple, Pragmatic Implementation.} We provide a lightweight CodeCarbon-based instrumentation layer for NVFlare that supports round-aware energy and CO$_2$e logging~\cite{mlco2_2025,NVIDIA25}.
  \item \textbf{Experimental validation on FL workloads.} We validate the approach on CIFAR-10 classification and multi-site retinal optic disk segmentation to demonstrate feasibility.
\end{itemize}

By offering a FL carbon tracking method with repeatable simulations, our work aims to strengthen the empirical foundations of green FL research and better align FL experimentation by enabling computing and communication systems that are measurably more sustainable and energy-aware.

In the remainder of this paper, we (1) define the carbon measurement protocol, (2) describe the NVFlare--CodeCarbon integration design and implementation, (3) report emissions and energy results for CIFAR-10 classification and retinal optic disk segmentation under controlled experimental conditions, and (4) discuss the outcomes of our FL scenarios alongside recommendations that standardize green FL.

\section{Methods}
\label{sec:methods}
Herein we present our carbon-accounting methodology for FL in NVFlare using CodeCarbon applied to simulated FL CIFAR-10 and retinal segmentation tasks, specify the communication-emissions model, and discuss performance scenarios for inefficient environments.

\subsection{Carbon Emission Measurement}
\label{sec:boundary}

Our experimental design aligns with prior green FL systems perspectives, emphasizing that the total footprint in FL is driven by both computation and communication, and by the operational reality of at-scale deployments~\cite{Qi24,yousefpour2023greenfederatedlearning}.
We measure emissions within a boundary designed to be comparable across FL deployments:
\begin{enumerate}
  \item \textbf{Client-side compute emissions (measured):} emissions attributable to the client training process, including CPU, GPU, and RAM contributions, recorded with CodeCarbon at process scope. This component dominates when training is compute-heavy and client hardware is heterogeneous.
  \item \textbf{Client-side non-training overhead (measured):} emissions during phases when the process is not performing gradient updates (e.g., waiting between rounds, evaluation, and one-time initialization).
  \item \textbf{Communication emissions (estimated):} emissions associated with transmitting model updates between server and clients. Because communication energy is not directly observable from the simulated training process through CodeCarbon, we estimate it from transmitted bytes and a configurable network energy-intensity model.
\end{enumerate}

In our NVFlare client script, we define explicit CodeCarbon ``tasks'' so that emissions can be attributed to phases consistently across experiments (\texttt{init}, \texttt{idle\_time}, \texttt{round\_k} for training, and \texttt{evaluate}) and we emit per-round logs to the FL server for aggregation.

\subsection{Federated Learning Experiments}
\label{sec:experiments}

\paragraph{Federated orchestration.}
We use NVFlare's job API to construct a FedAvg workflow with six clients, a fixed number of rounds, or task specific target accuracy. This configuration mirrors common FL evaluation settings while remaining lightweight enough for controlled emissions experiments. Conceptually, the server broadcasts the current global model each round, clients train locally, and the server aggregates client updates using FedAvg~\cite{mcmahan2023communicationefficientlearningdeepnetworks}.

\paragraph{Client-side emissions tracking.}
Each client instantiates a CodeCarbon \texttt{EmissionsTracker} configured for process-level measurement (\texttt{tracking\_mode="process"}) with power sampling every second (\texttt{measure\_power\_secs=1}). The tracker is parameterized by a geographic setting (\texttt{country\_iso\_code}, defaulting to \texttt{USA} in our jobs) to support region-consistent carbon-intensity assumptions and enable consistent comparisons across clients for this proof-of-concept~\cite{mlco2_2025}.

We structure explicit compute measurements using  CodeCarbon:
\begin{itemize}
  \item \textbf{Initialization (\texttt{init})} captures one-time startup overhead (framework initialization, dataset loading, first model setup).
  \item \textbf{Idle time (\texttt{idle\_time})} captures non-training periods per round (waiting/coordination).
  \item \textbf{Round compute (\texttt{round\_k})} wraps local training for round.
  \item \textbf{Train and Evaluate tasks} measure training and evaluation separately and record per-task duration for time-based analyses (train time vs.\ idle time).
\end{itemize}
For each task, CodeCarbon returns emissions and energy breakdowns (CPU/GPU/RAM), which we serialize into a per-round emissions\_data object attached to outgoing FLModel metadata. This follows prior FL systems work that reports client-side energy/emissions and aggregates at the server~\cite{yousefpour2023greenfederatedlearning}.

\paragraph{cuDNN initialization}
We observed an initial ``energy spike'' early in training attributable to cuDNN initialization and subsequent kernel autotuning, which is consistent with common deep-learning runtime behavior. To isolate this as a one-time cost, we separate the \texttt{init} task from subsequent rounds. Alternatively, it is possible to utilize a flag to disable cuDNN for controlled ablations when needed; however, this leads to increased emissions across all rounds and is therefore not recommended.

\subsection{CIFAR-10 Classification}
\label{sec:cifar}

We evaluate the measurement standard on CIFAR-10 with a compact CNN in PyTorch. The client script uses standard cross-entropy training with SGD and fixed hyperparameters (learning rate, momentum, batch size) and evaluates accuracy each round.

\paragraph{Non-IID partitioning.}
To emulate multi-site non-IID data, we partition CIFAR-10 into six client datasets using Dirichlet label sampling ($\alpha = 1.0$) with a fixed seed (\texttt{seed=0}) following common FL partitioning practice. The splitter writes per-site index files (\texttt{site-0.npy}, \dots), and each client loads its assigned indices and trains only on that subset.

\paragraph{Model and training loop.}
Clients train a small CNN (two convolution blocks plus three fully connected layers). Each round, a client receives global weights from NVFlare, loads them into the network, trains locally using SGD with momentum and cross-entropy loss, evaluates on the CIFAR-10 test set, and returns updated parameters plus metrics to the server.

\paragraph{Accuracy target.}
Our CIFAR-10 configuration reached $\sim$90\% test accuracy with 100 total epochs. The provided script performs one local epoch per round (10 total local epochs over 10 rounds); for the 100-epoch setting we increased local epochs while keeping the same orchestration and measurement hooks to meet the accuracy target.

\subsubsection{Retinal Optic Disk Segmentation}
\label{sec:retina}

We further validate the approach on a medical-imaging segmentation workload spanning five retinal sites (clients). Each client trains on its site-local images and labels; the server aggregates via FedAvg as in the CIFAR experiment. The target performance level is a 0.80 dice similarity coefficient accuracy threshold, chosen to reflect clinically relevant segmentation quality under site heterogeneity. The same emissions logging schema (\texttt{init}/training/\texttt{evaluate}/\texttt{idle\_time} plus communication estimate) is applied, enabling direct comparability of reporting across classification and medical segmentation workloads~\cite{yousefpour2023greenfederatedlearning}.

\subsection{Communication Emissions Estimation}
\label{sec:comm}

To incorporate communication in a reproducible way, we estimate per-round communication emissions using model-update size. Each client serializes the model \texttt{state\_dict} to compute the byte size of its transmitted update and reports this value to the server along with the per-round emissions payload.

On the server side, we extend NVFlare's FedAvg workflow to compute a communication-emissions estimate after aggregation. Our implementation converts bytes to gigabytes and applies (i) a configurable network energy intensity (default $I_{\mathrm{net}}=0.01$~kWh/GB) and (ii) a configurable grid emissions factor (default $F_{\mathrm{grid}}=0.475$~kg CO$_2$e/kWh), multiplying by 2 to reflect both downlink and uplink per round:
\[
E_{\mathrm{comm}} = 2 \cdot D_{\mathrm{GB}} \cdot I_{\mathrm{net}},
\qquad
C_{\mathrm{comm}} = E_{\mathrm{comm}} \cdot F_{\mathrm{grid}}.
\]
The controller accumulates total bytes across all clients/rounds, reports a run-level communication footprint, and exports both a pickle and a CSV summarizing per-round compute, communication, and idle metrics. This compute+communication boundary follows the FL sustainability literature's emphasis that communication can be a non-trivial portion of total footprint depending on model size, network, and number of rounds.

\subsection{Efficiency Tests (High / Medium / Low)}
\label{sec:tiers}

To evaluate whether our measurement standard is sensitive to client hardware efficiency and subsequent heterogeneous-client behavior, which is a key consideration in carbon-aware FL, we executed three controlled ``site efficiency tier'' settings by injecting deterministic and stochastic slowdowns into the client training loop:
\begin{enumerate}
  \item \textbf{High-efficiency site (baseline).} No artificial delays:
  \begin{itemize}
    \item No additional non-backpropagation iterations 
    \item No invoke of training loop sleep function
  \end{itemize}

  \item \textbf{Medium-efficiency site (compute-inefficient, minimal idle expected).}
  Adds extra forward-only work:
    \begin{itemize}
    \item 100 Additional non-backpropagation iterations
    \item Implemented as additional forward passes per batch (no backward/optimizer step), simulating reduced throughput while keeping timing relatively synchronized.
  \end{itemize}

  \item \textbf{Low-efficiency site (compute-inefficient, idle expected).}
  Adds extra forward-only work and per-step sleep:
  \begin{itemize}
    \item 100 Additional non-backpropagation iterations
    \item Invoke of training loop sleep function
    \item Sleep is sampled in ms per step from a Gaussian distribution (mean = 500); GPU synchronization is forced after extra forward-only iterations on CUDA, so the added work is reflected in measured time/energy.
  \end{itemize}
\end{enumerate}

Across all tiers, we log CPU/GPU/RAM energy, training time, and idle time, along with derived communication energy and CO$_2$e from model-update sizes. All tests were run under the same federated schedule (10 rounds; six clients) so that differences in measured compute emissions and round durations can be attributed to performance and idle behavior rather than changes in FL protocol. This design supports the paper's core objective: a repeatable, framework-integrated method to quantify how system-level inefficiencies (e.g., heterogeneous device performance) translate into measurable energy and carbon outcomes in green FL.

\section{Results}
All experiments were orchestrated using NVIDIA NVFlare's simulation workflow (\texttt{BaseFedJob} + \texttt{ScriptRunner}) with one GPU allocated to the simulator and multiple ``sites'' executed as simulated clients on the same node. Our primary runs used NVIDIA H100 and V100 hardware to test the portability of the measurement pipeline. Each experiment used six sites (CIFAR-10) or five sites (retinal optic disk segmentation) and ran until a target performance level was achieved.

\subsection{CIFAR-10 with Efficiency Testing}

\begin{figure}[t]
  \centering
  \includegraphics[width=\linewidth]{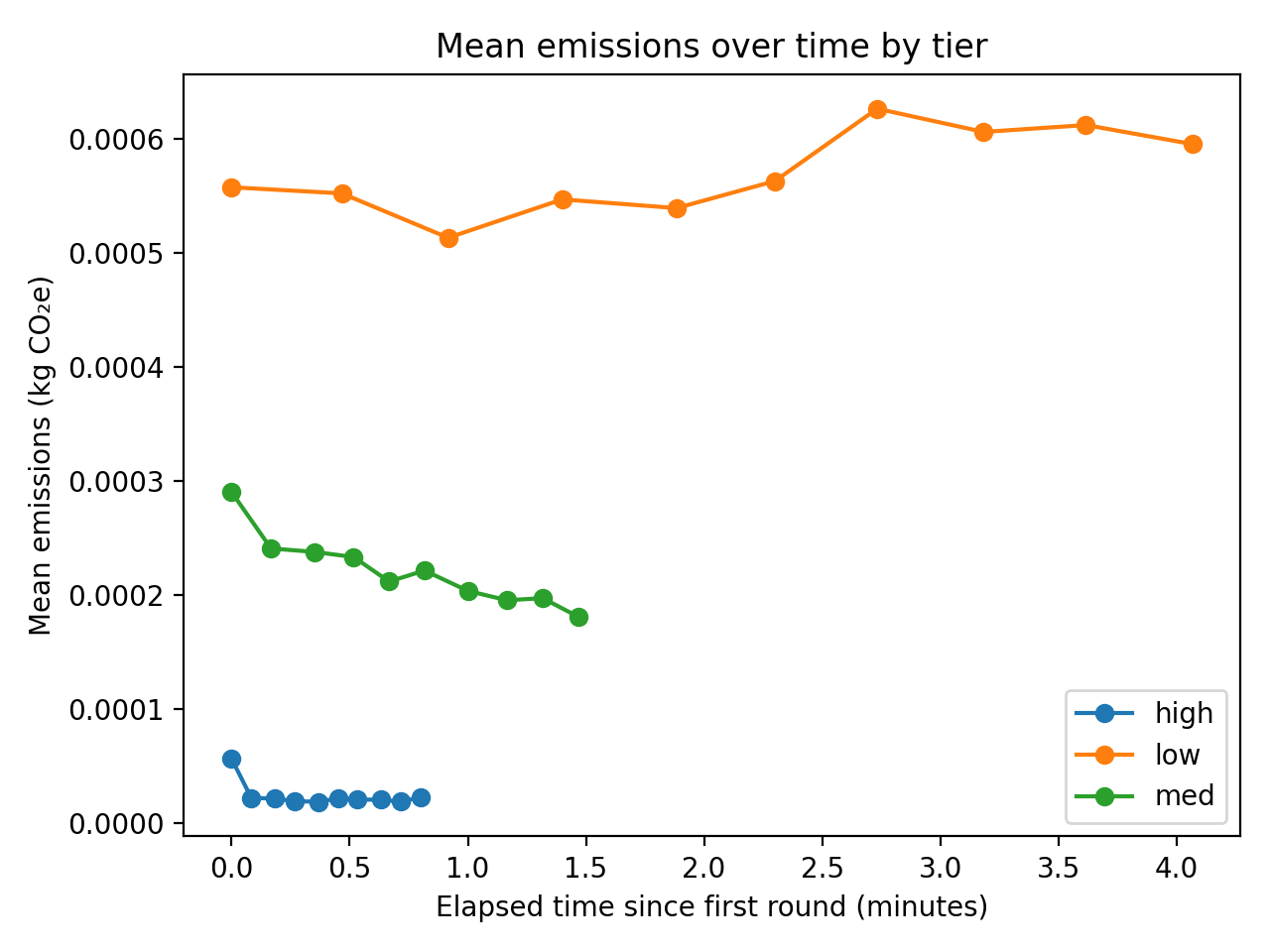}
  \caption{Mean emissions over time by client-performance efficiency tier (\textit{high}, \textit{medium}, \textit{low}) for the CIFAR-10 workload.}
  \label{fig:cifar_mean_emissions_over_time}
\end{figure}

We evaluated the proposed carbon-footprint measurement pipeline on a CIFAR-10 federated workload under three controlled client-efficiency scenarios (\textit{high}, \textit{medium}, \textit{low}) (see Section~\ref{sec:tiers}). All runs used the same FL protocol and identical instrumentation; only the injected slowdown parameters differed. Figure~\ref{fig:cifar_mean_emissions_over_time} summarizes the resulting emissions/energy profiles.

\paragraph{Runtime and overall footprint.}
Injected slowdowns produced a clear separation in both end-to-end wall-clock time and total carbon footprint. The \textit{high} baseline completed the 10-round run in $\approx 0.75$ minutes, whereas the \textit{medium} and \textit{low} regimes extended runtime to 1.52 and 4.23 minutes, respectively (Fig.~\ref{fig:cifar_mean_emissions_over_time}). This runtime inflation translated directly into higher total run-level emissions (compute + idle), which increased from $0.00161$ kgCO$_2$e in \textit{high} to $0.01343$ kgCO$_2$e in \textit{medium} and $0.03499$ kgCO$_2$e in \textit{low}. Relative to \textit{high} efficiency sites, these correspond to $8.34\times$ and $21.73\times$ increases under an otherwise identical FL protocol. Consistent with this trend, mean per client-round emissions rose from $0.000024$ kgCO$_2$e/round (\textit{high}) to $0.000221$ kgCO$_2$e/round (\textit{medium}) and $0.000571$ kgCO$_2$e/round (\textit{low}) (Table~\ref{tab:cifar_results_summary}). Mean energy per round similarly increased from $0.000062$ kWh/round (\textit{high}) to $0.000563$ kWh/round (\textit{medium}) and $0.001449$ kWh/round (\textit{low}).

\begin{table}[t]
  \centering
  \caption{CIFAR-10 run summary (6 clients, 10 rounds). Totals include measured client compute, measured idle bookkeeping, and estimated communication.}
  \label{tab:cifar_results_summary}
\resizebox{1.0\columnwidth}{!}{%
  \begin{tabular}{lcccc}
    \toprule
    Regime & Runtime & Mean energy & Mean CO$_2$e & Total CO$_2$e \\
     &(min) & (kWh/round) & (kg/round) & (kg) \\
    \midrule
    high   & 0.75 & 0.000062 & 0.000024 & 0.00161 \\
    medium & 1.52 & 0.000563 & 0.000221 & 0.01343 \\
    low    & 4.23 & 0.001449 & 0.000571 & 0.03499 \\
    \bottomrule
  \end{tabular}
  }
\end{table}

\paragraph{Communication vs.\ compute contributions.}
Estimated communication emissions were constant across resource environments, totaling $\approx 0.133$ gCO$_2$e per run. Thus, the fractional contribution of communication depended strongly on compute intensity: communication accounted for $\approx 8.3\%$ of total emissions in \textit{high}, but only $\approx 1.0\%$ in \textit{medium} and $\approx 0.4\%$ in \textit{low}, where training compute dominated. This highlights that a fixed communication baseline can become non-negligible only when the computation is highly efficient.

\FloatBarrier
\subsection{Retinal Optic Disk Segmentation}

\begin{figure}[t]
  \centering
  \includegraphics[width=\linewidth]{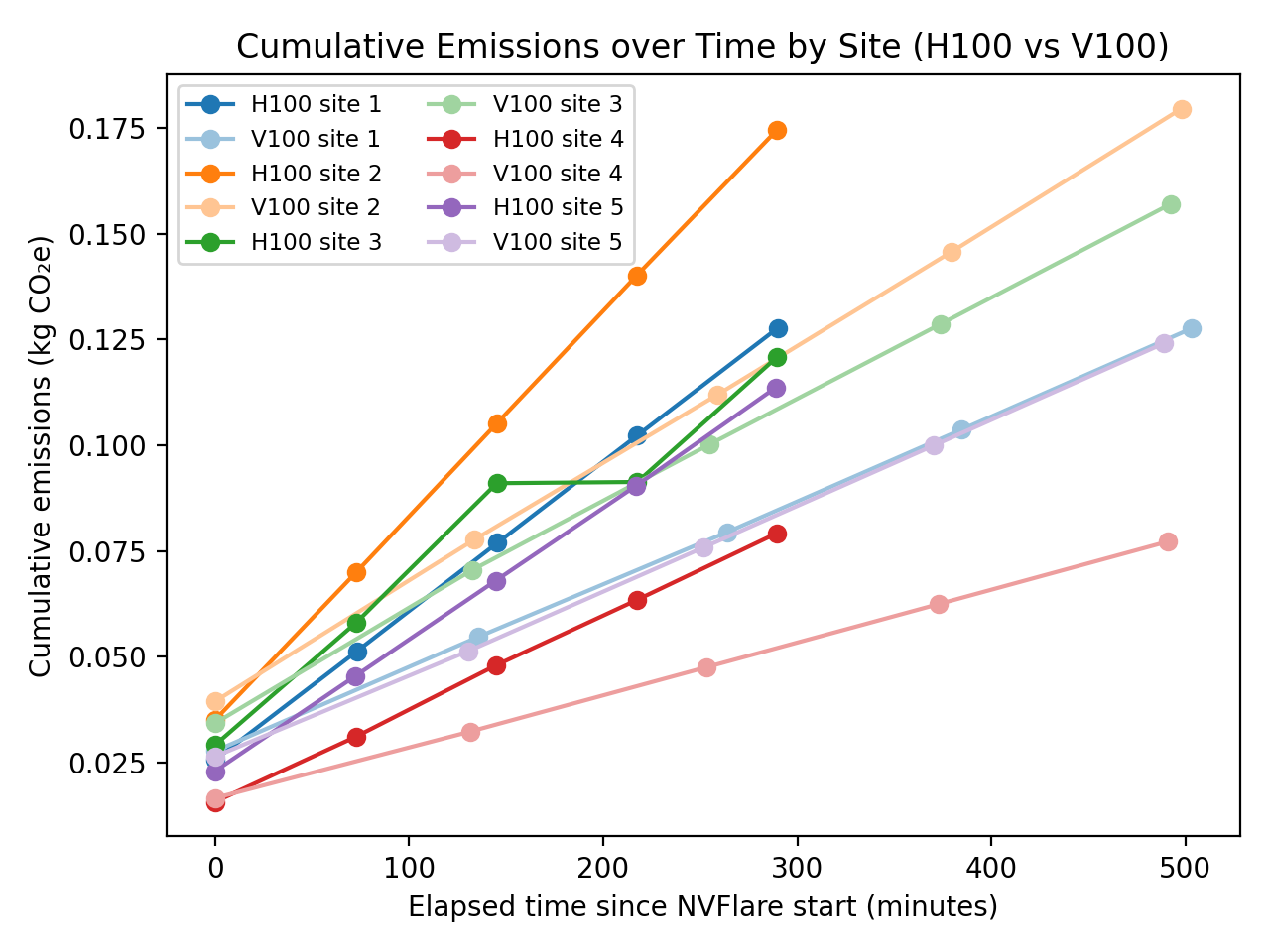}
  \caption{Cumulative (aggregate) CO$_2$e over time by site for the retinal segmentation task (H100 vs.\ V100).}
  \label{fig:site_agg_emissions_over_time}
\end{figure}

We evaluated our emissions accounting workflow on an FL scenario of a five-site retinal optic disk segmentation task executed under an identical protocol but with two different GPUs (H100 vs.\ V100). Figure~\ref{fig:site_agg_emissions_over_time} summarizes aggregate total emissions and total run time for each site with equivalent parameters and differences as a result of GPU type. Table~\ref{tab:retina_totals} summarizes the run-level totals read from our aggregation plots. The H100 tier remains relatively stable at approximately $0.065$--$0.068$~kWh/round. In contrast, the V100 tier exhibits higher energy in the early rounds (Round~1 $\approx 0.073$~kWh and Round~2 $\approx 0.072$~kWh), followed by a drop to $\approx 0.064$~kWh/round by Rounds~4--5. This pattern is consistent with a front-loaded overhead effect (e.g., initialization/autotuning or early-epoch compute intensity) that diminishes as the workflow proceeds.

\begin{table}[!t]
\caption{Retinal segmentation: run-level totals by site and comparative H100 (H) and V100 (V) GPUs.}
\label{tab:retina_totals}
\resizebox{1.0\columnwidth}{!}{%
\begin{tabular}{lcccccc}
\toprule
Site & CO$_2$e (H) & CO$_2$e (V) & Energy (H) & Energy (V) & Time (H) & Time (V) \\
 & (kg) & (kg) & (kWh) & (kWh) & (min) & (min) \\
\midrule
1 & 0.13 & 0.13 & 0.32 & 0.32 & 290.02 & 503.02 \\
2 & 0.17 & 0.18 & 0.44 & 0.46 & 289.65 & 497.88 \\
3 & 0.12 & 0.16 & 0.38 & 0.40 & 289.48 & 492.38 \\
4 & 0.08 & 0.08 & 0.20 & 0.20 & 289.27 & 490.62 \\
5 & 0.11 & 0.12 & 0.29 & 0.32 & 289.05 & 488.58 \\
\bottomrule
\end{tabular}
}
\end{table}

\paragraph{Run-level Site Variability.}
Figure~\ref{fig:site_agg_emissions_over_time} highlights notable site-to-site variability in total CO$_2$e, with totals spanning roughly $0.077$--$0.180$~kg CO$_2$e across sites. Site~2 is consistently the most carbon-intensive on both tiers ($\approx 0.175$~kg on H100 and $\approx 0.180$~kg on V100), whereas Site~4 is the least carbon-intensive ($\approx 0.079$~kg on H100 and $\approx 0.077$~kg on V100).

\section{Discussion}
Our results reinforce a key theme in green FL: emissions are not an intrinsic property of an algorithm alone, but the outcome of coupled decisions spanning model design, data and training protocols, system efficiency, orchestration, networking, and the carbon intensity of the energy supply. Across CIFAR-10 and retinal segmentation, four practical lessons emerge: (i) end-to-end carbon outcomes are highly sensitive to client-side inefficiency and variability even under an identical FL protocol, (ii) communication can appear dominant only in highly efficient regimes and depends strongly on the underlying network energy model, (iii) hardware tier differences may manifest primarily as time-to-train changes and site-specific effects rather than uniform reductions in total energy/CO$_2$e, and (iv) throughput (e.g., how many FL jobs can be executed concurrently per node) is primary determinant of carbon efficiency in practice.

\subsection{Efficiency Equates to Carbon Outcomes}

The CIFAR-10 experiments isolate performance heterogeneity as the primary experimental factor (same model, rounds, clients, and communication volume). Under this controlled setting, the \textit{medium} and \textit{low} efficiency scenarios increased total emissions by approximately $8.3\times$ and $21.7\times$ relative to \textit{high} (Table~\ref{tab:cifar_results_summary}). This separation demonstrates that ``green FL'' is not only an algorithmic property of aggregation or client selection, but also a function of how efficiently each participant executes the prescribed local workload. In other words, the same FL protocol can yield radically different carbon outcomes depending on throughput behavior; this is a practical concern that should be heavily considered for cross-silo deployments where clients differ in hardware, utilization, and background load.

A second implication is that per-round distributions matter. The CIFAR-10 results suggest an early-round ``startup'' overhead and regime-dependent variance, motivating carbon accounting that separates one-time initialization costs from steady-state training costs, and reporting that includes variability measures rather than only means. Such reporting is particularly important when comparing methods where the variance of per-round cost can be as consequential as the mean. Therefore, phase-aware logging is crucial: it enables attribution of emissions changes to training compute versus non-training overhead. Without phase separation, changes in end-to-end CO$_2$e can be misattributed to protocol design when they may instead reflect altered local execution efficiency or orchestration overhead.

More broadly, these behaviors motivate a standardized measurement boundary and logging schema. ``Emissions'' is not a single scalar property of FL; it reflects multiple coupled components (initialization, training compute, evaluation, idle/coordination, and communication) whose relative importance shifts across regimes and deployments. Without a fixed boundary and consistent per-round instrumentation, cross-paper comparisons are non-standard.

\subsection{Compute vs.\ Communication}

Our CIFAR-10 runs clarify when communication-focused optimizations are likely to matter. Because model size, number of rounds, and client participation were fixed, estimated communication emissions were approximately constant across sites ($\approx 0.133$ gCO$_2$e per run), so communication contributed a larger \emph{fraction} of emissions only when compute was highly efficient (\textit{high}). This highlights a common pitfall in green-FL evaluation: communication can be reported as a major contributor largely because the compute path is already optimized, whereas in inefficient sites, the dominant component is improving local execution efficiency and reducing idle/coordination waste.

At the same time, our communication estimate assumes a wired (data-center-like) energy intensity; in wireless settings (WiFi/LTE/5G), the communication component can be substantially larger and may dominate for small models or highly efficient clients. This aligns with prior observations that communication contributions are minimal under fiber-like assumptions but increase under mobile-networks, underscoring that communication should be considered through an explicit, reproducible modeling choices (intensity, system boundary, and uplink/downlink accounting)~\cite{GreenDFL2026,Yang21,Li21}.

\subsection{Hardware Tier Effects}

The retinal optic disk segmentation case study provides a complementary perspective: changing the GPU tier (H100 vs.\ V100) produced a large and consistent runtime separation across sites ($\sim$289--290 minutes vs.\ $\sim$489--503 minutes), yet total emissions were not uniformly reduced on the faster tier (Table~\ref{tab:retina_totals}). While some sites show near parity in total CO$_2$e across tiers (e.g., Site~1 and Site~4), others exhibit substantial increases on V100 (notably Site~3, $0.12 \rightarrow 0.16$~kg, and smaller increases for Sites~2 and~5). Importantly, energy totals track these outcomes closely: Site~3 also shows higher total energy on V100 ($0.38 \rightarrow 0.40$~kWh), whereas Sites~1 and~4 show effectively identical energy across tiers ($0.32$ and $0.20$~kWh, respectively). This indicates that ``faster'' hardware chiefly reduces wall-clock time, while total energy and CO$_2$e can remain similar or increase depending on site-level throughput and pipeline behavior.

These results motivate per-site and per-round reporting (not only totals) so that tier comparisons remain diagnostic and interpretable. Site-level summaries help disentangle whether differences arise from initialization overheads, data loading, mixed-precision behavior, or other pipeline-specific effects that may scale differently across hardware.

Importantly, per-run totals do not fully capture \emph{throughput} differences in practical simulation and shared-infrastructure settings. In our setup, an H100 node can sustain up to five concurrent NVFlare retinal segmentation jobs (i.e., 25 total sites), whereas a V100 node sustains only two (i.e., 10 total sites; thus, matching the throughput of five simultaneous FL runs (all 25 sites) would require the V100 to execute additional jobs, likely increasing total node energy due to prolonged runtime. Under this throughput-normalized view, the H100 becomes 2.5x more carbon-efficient per completed FL task. While such claims should be verified, the differential suggests that sustainable FL evaluation should report carbon normalized by delivered work (e.g., emissions per completed FL job, per site-round, or per accuracy target) when comparing hardware or system configurations.

\subsection{Carbon Intensity as a Major Consideration}

Carbon intensity (CI) is the primary determinant of emissions. The same workload can produce very different CO$_2$e outcomes depending on the electricity mix supplying the compute. Tools such as CodeCarbon make this factor explicit by associating energy use with geographically meaningful CI estimates, enabling comparable emissions calculations across sites worldwide. This also highlights a practical opportunity for carbon-aware orchestration: whenever privacy and governance constraints permit, carbon-intensive workloads can be preferentially scheduled to lower-CI locations or time windows, reducing total emissions without changing the underlying training algorithm.

\begin{figure}[t]
  \centering
  \includegraphics[width=\linewidth]{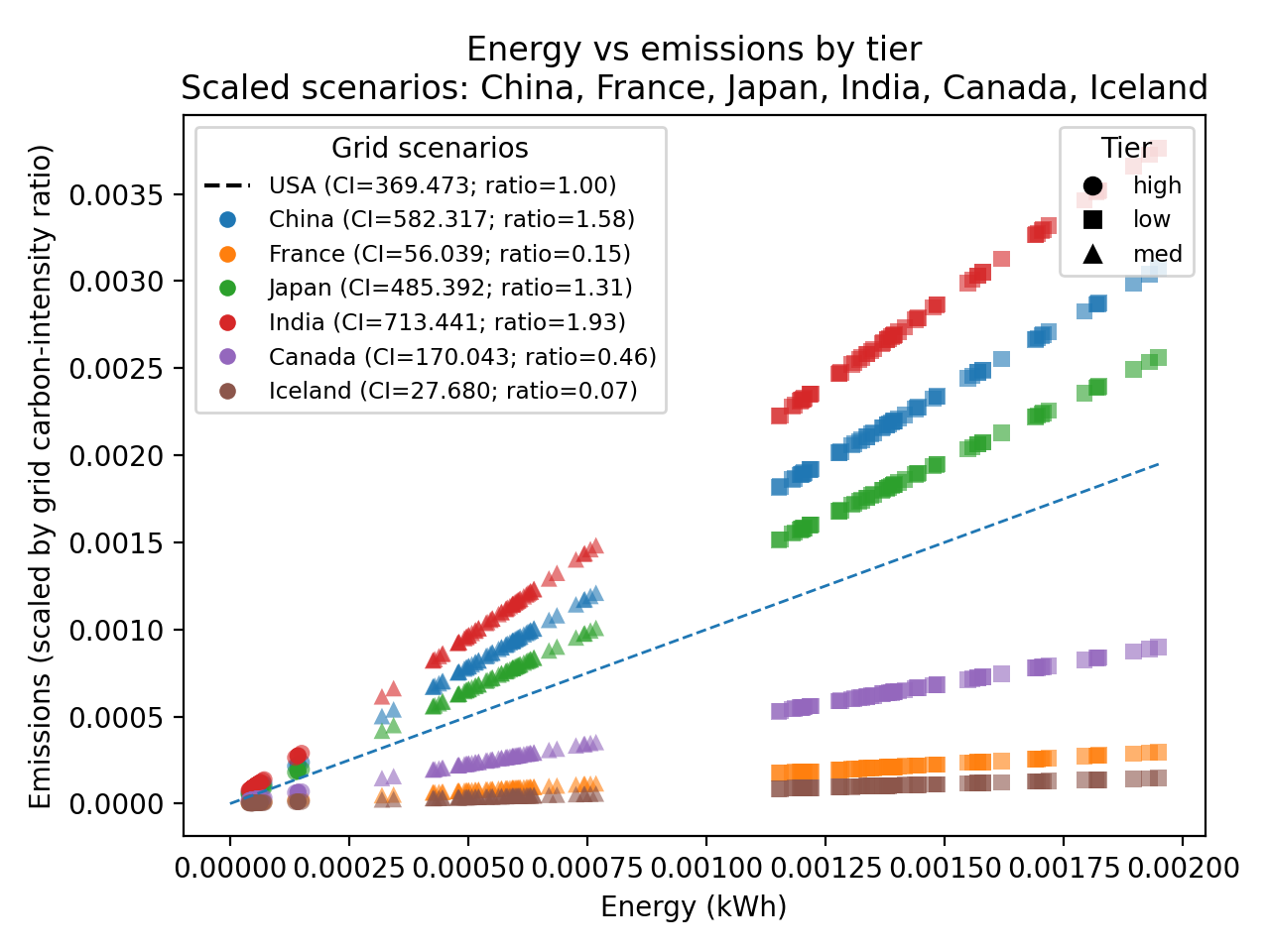}
  \caption{Energy vs.\ emissions under alternative grid carbon intensities (CI), illustrating how identical energy use leads to different CO$_2$e outcomes depending on location.}
  \label{fig:cifar_energy_vs_emissions}
\end{figure}

The energy--emissions relationship underscores that emissions are shaped not only by energy consumed, but by the carbon intensity of the electricity supplying that energy. When we rescaled emissions by grid CI, the same measured energy corresponded to dramatically different CO$_2$e outcomes: using the lowest and highest CI examples in Fig.~\ref{fig:cifar_energy_vs_emissions}, total run emissions for the \textit{low} regime span approximately $0.00001$--$0.0035$~kgCO$_2$e with equal energy consumption. This illustrates that \emph{where} training occurs can dominate emissions even when \emph{how much} energy is consumed is unchanged. As a result, energy-only reporting is insufficient for carbon-aware FL: evaluations should explicitly record the assumed region/carbon factor and, when possible, support time-varying or location-aware carbon accounting~\cite{Mehboob23}.

\begin{figure}[t]
  \centering
  \includegraphics[width=\linewidth]{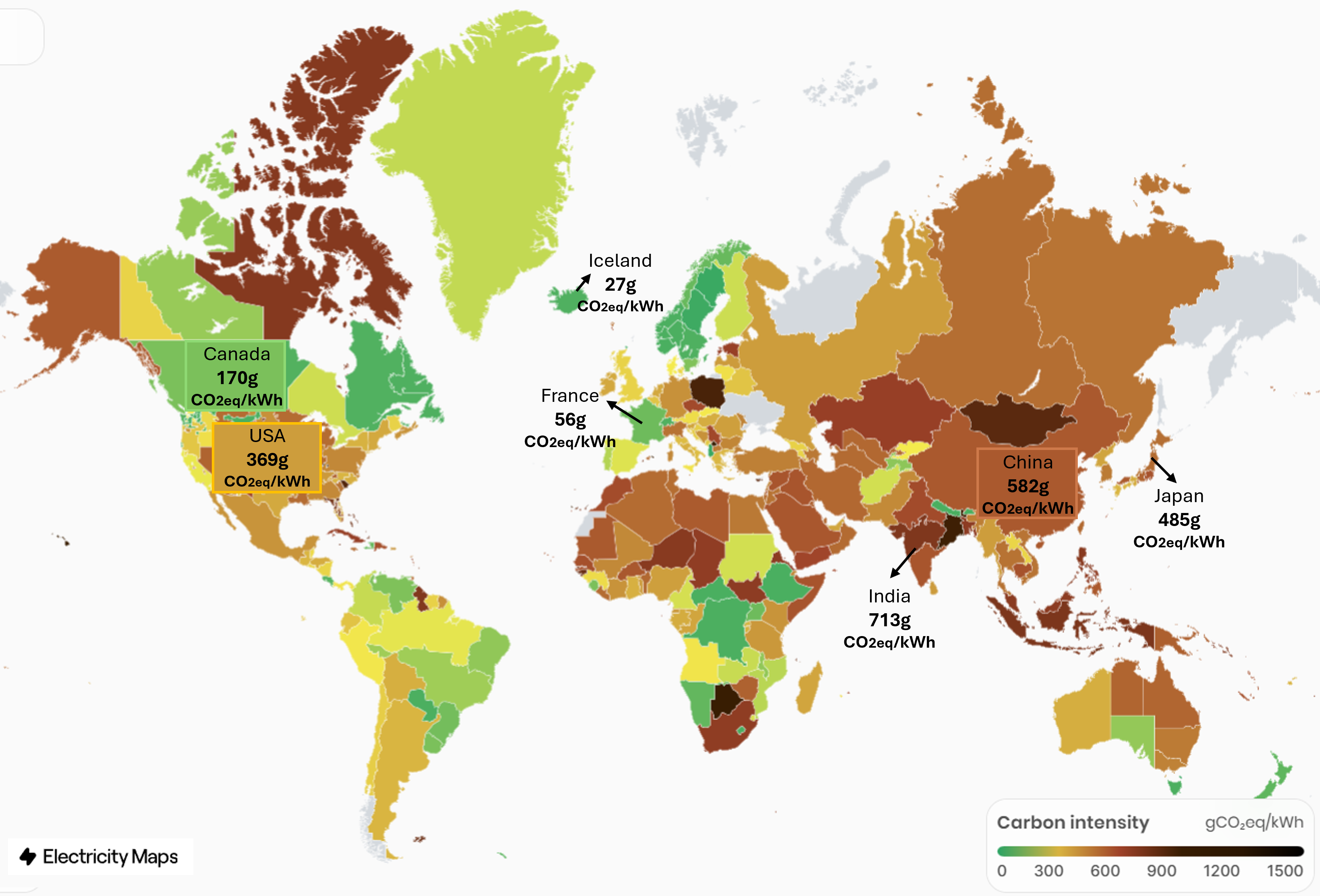}
  \caption{Global variation in grid carbon intensity, motivating location-aware and time-aware carbon accounting. Source: Electricity Maps live map.\protect\footnotemark}
  \label{fig:global_carbon}
\end{figure}
\footnotetext{\url{https://app.electricitymaps.com/map}}

\subsection{Standardized Carbon Tracking Benefits}

A key contribution of this work is operational: integrating standardized carbon-tracking instrumentation into a production-grade FL orchestrator. Using process-scoped tracking with explicit task boundaries allows emissions attribution to be consistent across experiments and platforms~\cite{mlco2_2025}. Embedding this within NVFlare’s execution model makes the approach portable for cross-silo and simulation settings, enabling researchers to run comparable experiments without rewriting measurement logic for each workflow~\cite{NVIDIA25}.

Tracking alone does not reduce emissions, but it enables principled optimization. By making per-round and per-phase costs visible, instrumentation supports explicit accuracy--carbon tradeoff analyses and enables fair comparison of algorithmic innovations (e.g., client scheduling, resource optimization, and device-aware policies) under a common accounting boundary~\cite{Hu22,Hsu22,Salh23,Kuswiradyo24}.

\subsection{Recommendations for Green FL}

Our experiments suggest several practical recommendations for conducting and reporting green FL studies:

\begin{enumerate}
  \item \textbf{Adopt and report clear measurement boundaries.}
  FL emissions are the sum of multiple components (client compute, client non-training overhead, orchestration/idle, and communication). Studies should state which components are \emph{measured} versus \emph{modeled}, and ensure the same boundary is used across compared methods (Section~\ref{sec:boundary}).

  \item \textbf{Report individualized metrics, not only totals.}
  Efficiency tests show that the same FL protocol can yield order-of-magnitude differences in CO$_2$e when client efficiency changes, and that early-run overheads can be visible and non-negligible. Reporting should include per-round curves and separate initialization reporting from steady-state training (Figures~\ref{fig:cifar_mean_emissions_over_time} and~\ref{fig:site_agg_emissions_over_time}).

  \item \textbf{Include site-level summaries to expose heterogeneity.}
  The retinal case study shows meaningful site-to-site variability and non-uniform tier effects (Table~\ref{tab:retina_totals}). Reporting only means or aggregates can hide worst-case sites that drive synchronization delays, idle time, or disproportionate carbon cost.

  \item \textbf{Report or estimate communication costs.}
  Communication emissions can appear dominant in highly efficient clients. Because the magnitude of this term depends strongly on network type (wired vs.\ WiFi/LTE/5G), the assumed intensity, and uplink/downlink accounting, green FL studies should publish these parameters and include simple sensitivity checks where feasible~\cite{GreenDFL2026,Yang21,Li21}.

  \item \textbf{Log carbon-intensity metadata alongside energy.}
  Our CI rescaling demonstrates that location can dominate carbon outcomes even when energy is unchanged (Figure~\ref{fig:cifar_energy_vs_emissions}). Evaluations should record the CI source (region, factor, and whether time-varying), enabling fair cross-site comparisons and supporting carbon-aware scheduling policies~\cite{Mehboob23}.
\end{enumerate}
\subsection{Limitations and Future Directions}

This work focuses on a practical, framework-integrated accounting layer rather than an exhaustive lifecycle assessment. Our communication component is model-based (bytes $\rightarrow$ energy $\rightarrow$ CO$_2$e) and thus depends on the chosen network-intensity parameterization; future work could integrate richer network telemetry or sensitivity analyses across realistic network types (e.g., WiFi, LTE/5G, data-center interconnect), while preserving reproducibility by keeping assumptions explicit.

Additionally, our current experiments use simulated clients on a single node for orchestration convenience; while the accounting approach is portable, real multi-node deployments may introduce additional orchestration and network effects that warrant further evaluation. A natural next step is to use this standardized instrumentation to evaluate carbon-aware FL control policies (e.g., client scheduling under carbon-intensity variation), since our logging already exposes the per-round structure needed to implement and validate such policies. 

\section{Conclusion}
This paper addressed a key barrier to reproducible green FL: the lack of a widely adopted, practical standard for measuring and reporting the carbon footprint of FL jobs across compute and communication. We introduced a lightweight, framework-integrated accounting approach that wraps CodeCarbon around NVFlare with explicit phase-aware tasks and round-level logging, and we complemented measured client compute and overhead with a transparent communication-emissions estimate derived from model-update payload sizes. Across CIFAR-10 and retinal optic disk segmentation, the proposed instrumentation revealed that (i) client-side inefficiency and variability can dominate total CO$_2$e under a fixed FL protocol, (ii) hardware tier differences may primarily shift wall-clock time without proportionally changing energy or emissions, and (iii) aggregate totals alone can obscure important per-site behaviors. Overall, utilizing CodeCarbon as a measurement wrapper for FL tasks orchestrated through NVFlare makes it easier to compare green FL methods under a consistent carbon accounting approach, strengthening the empirical foundations of the field, enabling direct comparisons across green FL methods, and establishing a practical foundation for future carbon-aware optimization and scheduling in federated systems. 

\bibliographystyle{ACM-Reference-Format}
\bibliography{base}

@String{Computing = "Computing" }

@String{Computer = "{IEEE} Computer" }

@Article{Yang21,
  author        = "Zhaohui Yang and Mingzhe Chen and Walid Saad and Chong Sang Hong and Mohammad Shikh‑Bahaei",
  title         = "Energy Efficient Federated Learning over Wireless Communication Networks",
  journal       = "IEEE Transactions on Wireless Communications",
  volume        = 20,
  number        = 3,
  month         = mar,
  year          = 2021,
  pages         = "1935--1949",
  doi           = "10.1109/TWC.2020.3037554",
  url           = "https://doi.org/10.1109/TWC.2020.3037554",
}

@inproceedings{Barbieri2023,
  author    = {Barbieri, Luca and Savazzi, Stefano and Kianoush, Sanaz and Nicoli, Monica and Serio, Luigi},
  title     = {A Carbon Tracking Model for Federated Learning: Impact of Quantization and Sparsification},
  booktitle = {2023 IEEE 28th International Workshop on Computer Aided Modeling and Design of Communication Links and Networks (CAMAD)},
  year      = {2023},
  month     = nov,
  pages     = {213--218},
  address   = {Edinburgh, United Kingdom},
  publisher = {IEEE},
  doi       = {10.1109/CAMAD59638.2023.10478391},
  isbn      = {979-8-3503-0350-6}
}

@inproceedings{Li21,
  author    = {Li, Peichun and Huang, Xumin and Pan, Miao and Yu, Rong},
  title     = {FedGreen: Federated Learning with Fine-Grained Gradient Compression for Green Mobile Edge Computing},
  booktitle = {2021 IEEE Global Communications Conference (GLOBECOM)},
  year      = {2021},
  month     = dec,
  pages     = {1--6},
  address   = {Madrid, Spain},
  publisher = {IEEE},
  doi       = {10.1109/GLOBECOM46510.2021.9685582}
}

@Article{Hu22,
  author        = "Yan Hu and Haibao Huang and Nuo Yu",
  title         = "Resource Optimization and Device Scheduling for Flexible Federated Edge Learning with Tradeoff Between Energy Consumption and Model Performance",
  journal       = "Mobile Networks \& Applications",
  volume        = 27,
  year          = 2022,
  pages         = "2118--2137",
  doi           = "10.1007/s11036-022-02009-2",
  url           = "https://doi.org/10.1007/s11036-022-02009-2",
}

@Article{Hsu22,
  author        = "Yuan‑Lin Hsu and Chia‑Feng Liu and Hung‑Yu Wei and Mehdi Bennis",
  title         = "Optimized Data Sampling and Energy Consumption in IIoT: A Federated Learning Approach",
  journal       = "IEEE Transactions on Communications",
  volume        = 70,
  number        = 12,
  month         = dec,
  year          = 2022,
  pages         = "7915--7931",
  doi           = "10.1109/TCOMM.2022.3216353",
  url           = "https://doi.org/10.1109/TCOMM.2022.3216353",
}

@Article{Salh23,
  author        = "Adeb Salh and Razali Ngah and Lukman Audah and Kwang Soon Kim and Qazwan Abdullah and Yahya M. Al‑Moliki and Khaled A. Aljaloud and Md. Hairul Nizam Talib",
  title         = "Energy‑Efficient Federated Learning with Resource Allocation for Green IoT Edge Intelligence in B5G",
  journal       = "IEEE Access",
  volume        = 11,
  year          = 2023,
  pages         = "16353--16367",
  doi           = "10.1109/ACCESS.2023.3244099",
  url           = "https://doi.org/10.1109/ACCESS.2023.3244099",
}

@Article{Albaseer24,
  author        = "Abdullatif Albaseer and Abegaz Mohammed Seid and Mohamed Abdallah and Ala Al‑Fuqaha and Aiman Erbad",
  title         = "Novel Approach for Curbing Unfair Energy Consumption and Biased Model in Federated Edge Learning",
  journal       = "IEEE Transactions on Green Communications and Networking",
  volume        = 8,
  number        = 2,
  month         = jun,
  year          = 2024,
  pages         = "865--877",
  doi           = "10.1109/TGCN.2024.3350735",
  url           = "https://doi.org/10.1109/TGCN.2024.3350735",
}

@Inproceedings{Fontenla24,
  author        = "Oscar Fontenla‑Romero and Blanca Guijarro‑Berdiñas and Elena Hernández‑Pereira and Blanca Pérez‑Sánchez",
  title         = "An Effective and Efficient Green Federated Learning Method for One‑Layer Neural Networks",
  booktitle     = "SAC'24: Proceedings of the 39th ACM/SIGAPP Symposium on Applied Computing",
  year          = 2024,
  pages         = "1050--1052",
  month         = apr,
  doi           = "10.1145/3605098.3636144",
  url           = "https://doi.org/10.1145/3605098.3636144",
  publisher     = "ACM",
  address       = "New York, NY, USA",
}

@Article{Kim24,
  author        = "Minsu Kim and Walid Saad and Mohammad Mozaffari and Mérouane Debbah",
  title         = "Green, Quantized Federated Learning over Wireless Networks: An Energy‑Efficient Design",
  journal       = "IEEE Transactions on Wireless Communications",
  volume        = 23,
  number        = 2,
  month         = feb,
  year          = 2024,
  pages         = "1386--1402",
  doi           = "10.1109/TWC.2023.3289177",
  url           = "https://doi.org/10.1109/TWC.2023.3289177",
}

@Article{Wang24,
  author        = "Jiali Wang and Yijie Mao and Ting Wang and Yuanming Shi",
  title         = "Green Federated Learning over Cloud‑RAN with Limited Fronthaul Capacity and Quantized Neural Networks",
  journal       = "IEEE Transactions on Wireless Communications",
  volume        = 23,
  number        = 5,
  month         = may,
  year          = 2024,
  pages         = "4300--4314",
  doi           = "10.1109/TWC.2023.3317129",
  url           = "https://doi.org/10.1109/TWC.2023.3317129",
}

@Article{Kuswiradyo24,
  author        = "Primatar Kuswiradyo and Binayak Kar and Shan‑Hsiang Shen",
  title         = "Optimizing the Energy Consumption in Three‑Tier Cloud–Edge–Fog Federated Systems with Omnidirectional Offloading",
  journal       = "Computer Networks",
  volume        = 250,
  year          = 2024,
  articleno     = 110578,
  numpages      = 1,
  doi           = "10.1016/j.comnet.2024.110578",
  url           = "https://doi.org/10.1016/j.comnet.2024.110578",
}

@Article{Qi24,
  author        = "Yuanhang Qi and M. Shamim Hossain",
  title         = "Harnessing Federated Generative Learning for Green and Sustainable Internet of Things",
  journal       = "Journal of Network and Computer Applications",
  volume        = 222,
  year          = 2024,
  articleno     = 103812,
  numpages      = 1,
  doi           = "10.1016/j.jnca.2023.103812",
  url           = "https://doi.org/10.1016/j.jnca.2023.103812",
}

@Misc{Savazzi22,
  author        = "Raffaele Savazzi and Leo Loven and Allan B. Poulsen",
  title         = "Energy and Carbon Footprint Analysis of Distributed and Federated Learning: An Overview and Experimental Evaluation",
  year          = 2022,
  eprint        = "2209.00050",
  archivePrefix = "arXiv",
  primaryClass  = "cs.LG",
  url           = "https://arxiv.org/abs/2209.00050",
}

@Article{Ciroi23,
  author        = "Xinchi Qiu and Titouan Parcollet and Javier Fernandez‑Marques and Pedro P.B. Gusmao and Yan Gao and Daniel J. Beutel and Taner Topal and Akhil Mathur and Nicholas D. Lane",
  title         = "A First Look into the Carbon Footprint of Federated Learning",
  journal       = "Journal of Machine Learning Research",
  volume        = 24,
  number        = 129,
  year          = 2023,
  pages         = "1--23",
  url           = "http://jmlr.org/papers/v24/21-0445.html",
}

@Misc{Mehboob23,
  author        = "Talha Mehboob and Noman Bashir and Jesus Omana Iglesias and Michael Zink and David Irwin",
  title         = "EcoLearn: Optimizing the Carbon Footprint of Federated Learning",
  year          = 2023,
  eprint        = "2310.17972",
  archivePrefix = "arXiv",
  primaryClass  = "cs.LG",
  url           = "https://arxiv.org/abs/2310.17972",
  note          = "arXiv preprint, version updated in 2025",
}

@Misc{mlco2_2025,
  author        = "mlco2",
  title         = "CodeCarbon: Track Emissions from Compute and Recommend Ways to Reduce Their Impact on the Environment",
  year          = 2025,
  howpublished  = "Software repository",
  url           = "https://github.com/mlco2/codecarbon",
  note          = "Accessed 2025",
}

@Misc{NVIDIA25,
  author        = "NVIDIA",
  title         = "NVFlare: Federated Learning from Simulation to Real World",
  year          = 2025,
  howpublished  = "Software framework and documentation",
  url           = "https://github.com/NVIDIA/NVFlare",
  note          = "Accessed 2025",
}

@misc{mcmahan2023communicationefficientlearningdeepnetworks,
      title={Communication-Efficient Learning of Deep Networks from Decentralized Data}, 
      author={H. Brendan McMahan and Eider Moore and Daniel Ramage and Seth Hampson and Blaise Agüera y Arcas},
      year={2023},
      eprint={1602.05629},
      archivePrefix={arXiv},
      primaryClass={cs.LG},
      url={https://arxiv.org/abs/1602.05629}, 
}

@article{BOLONCANEDO2024128096,
title = {A review of green artificial intelligence: Towards a more sustainable future},
journal = {Neurocomputing},
volume = {599},
pages = {128096},
year = {2024},
issn = {0925-2312},
doi = {https://doi.org/10.1016/j.neucom.2024.128096},
url = {https://www.sciencedirect.com/science/article/pii/S0925231224008671},
author = {Verónica Bolón-Canedo and Laura Morán-Fernández and Brais Cancela and Amparo Alonso-Betanzos}}

@misc{yousefpour2023greenfederatedlearning,
      title={Green Federated Learning}, 
      author={Ashkan Yousefpour and Shen Guo and Ashish Shenoy and Sayan Ghosh and Pierre Stock and Kiwan Maeng and Schalk-Willem Krüger and Michael Rabbat and Carole-Jean Wu and Ilya Mironov},
      year={2023},
      eprint={2303.14604},
      archivePrefix={arXiv},
      primaryClass={cs.LG},
      url={https://arxiv.org/abs/2303.14604}, 
}

@misc{arputharaj2025greenfederatedlearningcarbonaware,
      title={Green Federated Learning via Carbon-Aware Client and Time Slot Scheduling}, 
      author={Daniel Richards Arputharaj and Charlotte Rodriguez and Angelo Rodio and Giovanni Neglia},
      year={2025},
      eprint={2509.08980},
      archivePrefix={arXiv},
      primaryClass={cs.LG},
      url={https://arxiv.org/abs/2509.08980}, 
}

@inproceedings{FedSynthesis2026,
author = {Cantali, G\"{o}kcan and G\"{u}r, G\"{u}rkan and Stiller, Burkhard},
title = {FedSynthesis: A Flower-based Framework for Carbon-Reduced Federated Learning},
year = {2026},
isbn = {9798400722851},
publisher = {Association for Computing Machinery},
address = {New York, NY, USA},
url = {https://doi.org/10.1145/3773274.3774265},
doi = {10.1145/3773274.3774265},
booktitle = {Proceedings of the 18th IEEE/ACM International Conference on Utility and Cloud Computing},
articleno = {10},
numpages = {9},
series = {UCC '25}
}

@article{Qui2023carbonfootprint,
author = {Qiu, Xinchi and Parcollet, Titouan and Fernandez-Marques, Javier and Gusmao, Pedro P. B. and Gao, Yan and Beutel, Daniel J. and Topal, Taner and Mathur, Akhil and Lane, Nicholas D.},
title = {A first look into the carbon footprint of federated learning},
year = {2023},
issue_date = {January 2023},
publisher = {JMLR.org},
volume = {24},
number = {1},
issn = {1532-4435},
journal = {J. Mach. Learn. Res.},
month = jan,
articleno = {129},
numpages = {23}
}

@article{GreenDFL2026,
title = {GreenDFL: A framework for assessing the sustainability of Decentralized Federated Learning systems},
journal = {Information and Software Technology},
volume = {190},
pages = {107937},
year = {2026},
issn = {0950-5849},
doi = {https://doi.org/10.1016/j.infsof.2025.107937},
url = {https://www.sciencedirect.com/science/article/pii/S0950584925002769},
author = {Chao Feng and Alberto {Huertas Celdrán} and Xi Cheng and Gérôme Bovet and Burkhard Stiller}
}

@misc{Flower2022,
      title={Flower: A Friendly Federated Learning Research Framework}, 
      author={Daniel J. Beutel and Taner Topal and Akhil Mathur and Xinchi Qiu and Javier Fernandez-Marques and Yan Gao and Lorenzo Sani and Kwing Hei Li and Titouan Parcollet and Pedro Porto Buarque de Gusmão and Nicholas D. Lane},
      year={2022},
      eprint={2007.14390},
      archivePrefix={arXiv},
      primaryClass={cs.LG},
      url={https://arxiv.org/abs/2007.14390}, 
}

@article{Thakur2025GreenFL,
author = {Thakur, Dipanwita and Guzzo, Antonella and Fortino, Giancarlo and Piccialli, Francesco},
title = {Green Federated Learning: A New Era of Green Aware AI},
year = {2025},
issue_date = {August 2025},
publisher = {Association for Computing Machinery},
address = {New York, NY, USA},
volume = {57},
number = {8},
issn = {0360-0300},
url = {https://doi.org/10.1145/3718363},
doi = {10.1145/3718363},
journal = {ACM Comput. Surv.},
month = mar,
articleno = {194},
numpages = {36}
}

\end{document}